# Electrostatic Gating of Ionic Conductance Through Heterogeneous van der Waals Nanopores

Aaron H. Barajas-Aguilar,[1,#] Matthew Schiel,[1,#] Ethan Cao[1], DaVante Cain[1], Margaret L. Berrens,[2] Fikret Aydin,[2] Tuan Anh Pham,[2] Javier Sanchez-Yamagishi[1,*], Zuzanna S. Siwy[1,*]

[1]Department of Physics and Astronomy, University of California, Irvine, Irvine, CA 92697

[2]Quantum Simulations Group and Laboratory for Energy Applications for the Future, Lawrence Livermore National Laboratory, Livermore, California 94551, USA

## Abstract

Nanofluidic ionic transistors typically require gate voltages above 1 V and operate only at sub-millimolar ionic strengths, limiting their biocompatible applications. We demonstrate ionic transistors consisting of single sub-10 nm nanopores drilled in van der Waals (vdW) heterostructures with internal gate electrodes made of few-layer graphene. These devices deliver up to 10-fold current modulation at gate voltages as low as 0.3 V in 10 mM KCl, and ~2-fold modulation at near physiological 100 mM KCl. Baseline conductance with no gate shows surface-charge-dominated transport below 100 mM KCl consistent with negatively charged hBN walls and ~5 nm opening of the pores. The surface-charge and the electrochemical asymmetry introduced by the three-electrode configuration govern the device's behavior: negative gate voltage ($V_G$) enriches ionic concentrations and enhances current, whereas positive $V_G$ induces a local depletion zone that suppresses transport. The current modulation by $V_G$ is dependent on the polarity of the transmembrane potential and leads to ion current rectification. Molecular dynamics simulations of a nanopore in a hBN-graphene-hBN stack reveal confinement and surface charge dependent suppression of relative permittivity of interfacial water. Continuum modeling with radially varying interfacial water permittivity reproduces the asymmetric I–V characteristics and explains how the embedded gate sculpts local potential and ion concentrations. By enabling sub 0.5 V control of ionic transport at up to 100 mM salt concentrations, these devices address a key barrier in nanofluidics and open the pathway to low power ionic circuits and biosensing.

[#]These Authors contributed equally

[*] Corresponding Authors: javier.sanchezyamagishi@uci.edu, zsiwy@uci.edu

Ion transport at the nanoscale exhibits phenomena not observed in microstructures due to the large surface to volume ratio of nanopores and nanochannels that makes the transport extremely sensitive to the electrochemical properties of the pore walls.[1-3] Changing the polarity and magnitude of surface charges tunes ion selectivity and conductance of the structures. Introducing surface charge patterns can break electrochemical symmetry of nanopores leading to rectification and diode-like characteristics of the transmembrane current,[4-6] behavior observed in many biological channels in a cell membrane.[7]

The field of nanofluidics has also been inspired by gating achieved in electronic solid-state devices, specifically the ability to control ionic current by integration of electrically addressable gates.[8, 9] If we could electrically, on demand, change the local electric potential in the pore, we could create ionic equivalents of electronic devices and connect them into circuits to achieve logic gates and ionic amplifiers. [10-12]

An ionic transistor modulates ion current by means of an electrically addressable electrode, and several examples of such systems have already been reported in the literature.[10] Sub-10 nm in diameter nanopore field-effect transistors, fabricated using electron beam lithography and atomic layer deposition, provided gating in KCl concentrations of 1 mM and lower.[13] At a gate voltage of -1 V and at 0.1 mM KCl, the transmembrane ion current was enhanced by a factor of up to 10. Similarly, 20 nm thick nanofluidic channels with two independent gates functioned as reconfigurable ionic diodes at mM concentrations of KCl and gate voltages of ~1 V,[14] while larger nanofluidic channels required gate voltages of tens of V to achieve ion current modulation.[15] Ionic ambipolar transistors were reported as well, where carbon nanotubes templated into alumina were biased by an integrated gate.[16] The smallest opening of the carbon nanotubes was 5 nm in diameter. When gate voltages of -1 V and +1 V were applied to the membrane, two conductance states were obtained with an on-off ratio of up to 16,000 in 0.1 mM KCl. Another type of ionic transistors utilized atomic-scale channels in graphene oxide with interlayer spacing of ~0.45 nm.[17] Applying gate voltage to the material changed the energy barrier for ion intercalation into the channel and enabled the transport for negative gate voltages exceeding -0.8 V, while blocking the transport for positive gate voltages.

Ionic transistors based on channels with an opening significantly larger than the Debye screening length were reported as well. A voltage of 1 V at a gate placed on top of a membrane with a 300 nm pore increased surface charge density on the pore walls and ion selectivity, without significantly influencing the pore conductance. [18] Even larger scale ionic transistors were prepared based on microchannels and ion exchange membranes offering high on-off ratios at the cost of high gate potentials of few volts. [19]

Nearly all ionic transistors reported thus far required gate voltages of at least 1 V to achieve significant modulations of ion current, and gating was observed in KCl concentrations of ~ 1 mM or lower. Reducing the magnitude of the gate voltage while being able to modulate the ionic current at physiological salt concentrations of 100 mM is, however, important for future applications in ionic circuits, biocompatible devices, and biosensing.[16] Low gate voltages would also mitigate energy requirements for operation, minimize undesired electrochemical reactions, and are expected to increase the longevity of the nanopore systems.

Van der Waals (vdW) materials, such as graphite or hexagonal boron nitride and others, offer a unique host for nanopore based devices.[20-24] Due to their layered structure, vdW materials can be stacked to form heterogenous membranes[25,26-32] with atomically sharp interfaces between materials that exhibit vastly different properties. In principle, this enables the design of electric fields within a nanopore at the nanoscale, which is difficult to achieve with conventional nanofabrication techniques. Moreover, the single-crystal nature of vdW materials departs strongly from the amorphous or granular nature of typical substrates used in solid-state nanopores.

Here, we utilize the unique properties of vdW films and report the design and performance of atomically-precise ionic transistors that function at high salt concentrations, operate at gate voltages as low as 0.3 V, and offer modulation factors of up to 10x in 10 mM KCl. Our ionic transistor consists of a ~5 nm diameter nanopore drilled in a vdW heterostructure comprising two layers of hBN encapsulating an internal graphene layer that acts as the gate electrode. As we and others have shown, transport properties of pores with aspect ratio significantly larger than 1 are much more sensitive to the properties of the pore walls than short pores whose conductance is mostly determined by the region at the pore entrance.[33,34-36] The thickness of the hBN layers was therefore tuned to ensure that the ionic transistor nanopores had an aspect ratio of at least 5. The multilayer hBN and graphene in our devices are therefore in contrast to single-layer hBN, graphene, and $MoS_2$ films used for single molecule detection.[20, 21, 34-38] The sub-10 nm opening diameter and large aspect ratio as well as proximity of the gate electrode to the solution of our nanopores enabled achieving gating of ion currents in physiological conditions and sub-0.5V gate voltages.

## Results and Discussion

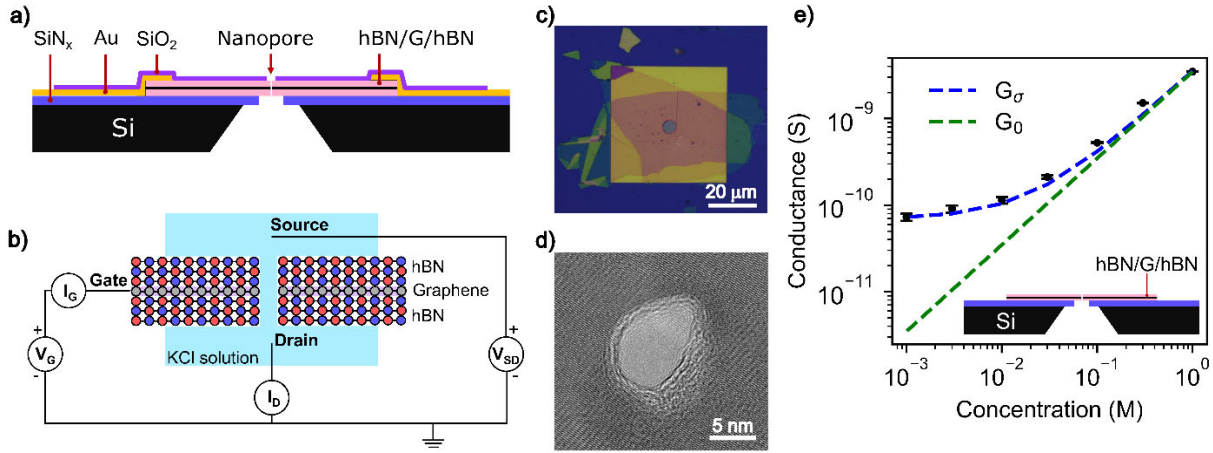

**Figure 1**. **Nanopore ionic transistor in heterogeneous stacks of van der Waals (vdW) materials with internal graphene gate electrode**. *a) Cross-section schematic of the assembled stack of graphene and two layers of hBN. b) Measurement circuit consisting of two independent voltages: source-drain, $V_{SD}$, and gate-drain, $V_G$. c) Optical image of the assembled stack. d) Transmission electron microscope (TEM) image of the nanopore drilled by electron beam (Device A). e) Ionic conductance, G, versus KCl concentration for the nanopore shown in (d) at pH=6. $G_\sigma$ and $G_0$ are the fits of the experimental data with Eq. (1)[39] when considering -2.3 mC/m$^2$ and zero effective surface charge of the pore walls, respectively. For this sample, the top and bottom hBN layer thicknesses were 16.7 nm and 13.7 nm, respectively, and the middle graphene layer was 1 nm thick.*

A scheme of the ionic transistor we designed is shown in Figure 1 a, b. The vdW heterostructure consists of a sequential stack of hBN, few-layer graphene, and another hBN, which are assembled following a dry transfer technique.[27] The final stack is transferred onto a TEM grid containing a 200 nm thick SiN$_x$ membrane with a 5 μm diameter pore (Norcada Industrial Company), Figure 1c. Gold electrodes were attached to the graphene gate using electron beam lithography and metal evaporation following a liftoff process (see the Methods section for details). To avoid direct exposure to the solution, the metallic contacts were passivated by a 170 nm thick layer of SiO$_2$ deposited using plasma enhanced chemical vapor deposition. A window on the SiO$_2$ layer was subsequently etched to expose the vdW membrane while leaving the electrodes protected. A nanopore was drilled into the vdW membrane with a focused electron beam in a transmission electron microscope (TEM) (Figure 1d).[40, 41] This protocol allowed us to consistently fabricate sub-10 nm in diameter nanopore transistors within the vdW membranes.

To characterize the conductance of the ionic transistor under gating, we measured the samples in KCl solutions at concentrations between 0.1 mM and 1 M. A custom electrolytic cell with an incorporated PCB board allowed us to establish electrical connection to the graphene gate electrode (see Supporting Information Section 1). The measurements entailed application of two independent voltages: the source-drain voltage, $V_{SD}$, which drives the ions through the nanopore, and the gate voltage, $V_G$, which modulates the local ion concentration in the nanopore (Figure 1b).

**Conductance measurements without gating reveal surface charge dominated transport through vdW nanopores.**

Nanopores prepared in the vdW heterostructures were first characterized without any gate modulation. Under the application of only $V_{SD}$, we observed that vdW nanopores behaved similarly to other nanopore systems (Figure 1e).[42] Namely, the nanopore conductance followed a linear dependence on the salt concentration until a threshold concentration of 100 mM KCl, and leveled off at lower concentrations. Due to the effective screening of surface charge at 1 M KCl,[1, 41] we used the conductance at this concentration to calculate the pore diameter. In the calculations we assumed the pore was cylindrical in shape with length, L = 31 nm (determined by AFM measurements of the stack), and filled with a solution whose conductivity was equal to the bulk conductivity of 1 M KCl, i.e. 10 S/m. For device A shown in Figure 1, the electrochemically measured diameter was ~4 nm, which is in good agreement with the 6 nm diameter measured by TEM. The small discrepancy between these two values might stem from possible double-conical shape of pores drilled in TEM, as shown before for nanopores in silicon nitride films.[43, 44]

The plateauing of conductance at ~100 mM KCl stems from the sub-10 nm opening and the presence of negative surface charges on hBN, shown before to enhance ionic transport by adding surface conductance.[1, 42] Once the thickness of the screening length approaches the pore radius, the ionic transport of the system is dominated by the surface conductance, i.e. the number of ions in the pore is determined by the surface charge density, not the bulk salt concentration. The magnitude of the residual conductance at low concentrations can be used to determine the effective surface charge density of the pore, using eq. (1).[39, 45]

$$G = (\frac{\pi d^2}{4L})[(\mu_K + \mu_{Cl})n_{KCl}e + \mu_K \frac{4\sigma}{d}] \tag{1}$$

Where d and L, are pore diameter and length, respectively, and σ is the surface charge density on the pore walls, and $\mu_K$ and $\mu_{Cl}$ are the mobilities of the potassium and chloride ions, respectively.

At pH 6, the surface charge density obtained from Figure 1e is -2.3 mC/m$^2$, which falls within the wide range of previously reported values for surface charge density of hBN surface and hBN nanopores of -1.2 and -250 mC/m$^2$.[46-49] Note that the leveling off of the currents already in a high concentration of 100 mM KCl (Figure 1e) supports the nanoscopic opening of the pore as well as the dominant role of the internal surface charges for the properties of ionic transport. We measured similar behavior in other vdW nanopores (Figure S2), and calculated surface charge densities in the same order of magnitude.

The presence of negative surface charges on the edges of the hBN nanopore was further verified by conductance measurements of the hBN-graphene-hBN pore as a function of solution pH (see Supporting Information, section 2). As the solution becomes more acidic, the residual conductance at low concentrations decreases, indicating a lower effective negative surface charge. These observations are in agreement with recent studies that revealed adsorption of OH$^-$ groups as the primary source of the effective surface charge on hBN. [48] The lower availability of OH$^-$ groups with lower pH, reduces the surface charge and, consequently, reduces the conductance at low salt concentrations.

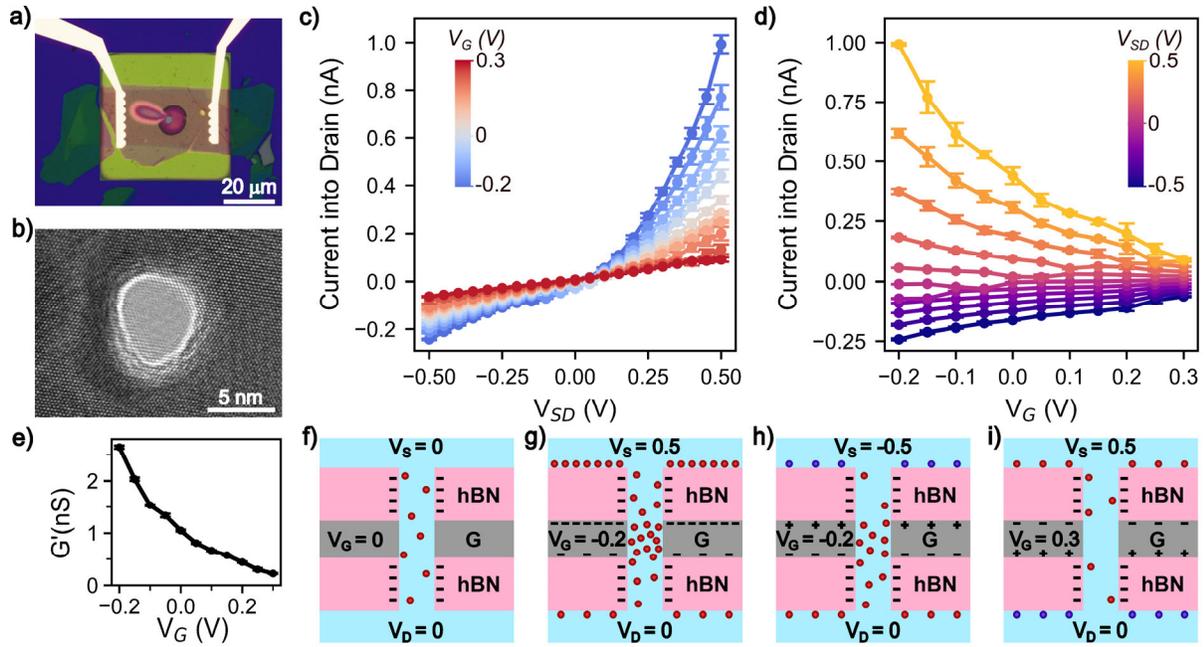

**Figure 2. Performance of vdW ionic transistors with gate voltage**. *a) Optical image of Device B with layer thicknesses of 30 nm, 2.5 nm, and 6 nm for the top hBN, graphene, and bottom hBN respectively, leading to a total membrane thickness of 38.5 nm. b) TEM image of the 5 nm in diameter nanopore in Device B. c) Drain current, $I_D$, vs source-drain voltage, $V_{SD}$, curves at different gate voltages, $V_G$, in 10 mM KCl for Device B. The results show the ionic current through the nanopore is modulated by $V_G$, such that $I_D$ increases (decreases) for negative (positive) $V_G$ in comparison with the $V_G$ = 0 V case. The ion current modulation is larger for positive $V_{SD}$ values. d) Plot of ion currents from panel (c) versus $V_G$ reveals that for $V_{SD}$ = 0.5 V there is a ~10 times modulation of the current for the $V_G$ range measured (0.3V to –0.2V). e) Local conductance (G') as a function of $V_G$, where G'=$\Delta I_D/\Delta V_{SD}$ was calculated in the range of $V_{SD}$ from 0.2 to 0.4 V. The local conductance also undergoes 10 times modulation in the same range of applied gate voltages. f), g), h, and i) show schemes of local ionic concentrations for the most representative cases: (f) $V_{SD}$ = 0 and $V_G$ = 0, g) $V_{SD}$ > 0, $V_G$ < 0, h) $V_{SD}$ < 0, $V_G$ < 0 and, (i) $V_{SD}$ > 0, $V_G$ > 0. Red and blue circles represent positive and negative ions respectively.*

**vdW Ionic transistors gate ion current with sub-0.5 V gate voltages.**

After completing test experiments with nanopores in the hBN-graphene-hBN stacks without electrodes, we performed electrochemical measurements of devices containing nanofabricated contacts (Figure 2a,b). We expected the transmembrane current, $I_D$, to be gated with $V_G$ significantly lower than 1 V, as shown before with graphene nanopore transistors whose opening was diminished by the presence of nanobubbles.[50]

Figure 2c, d shows results for device B, that was subjected to source-drain voltage, $V_{SD}$, from -0.5 to 0.5 V, and gate voltage, $V_G$, from -0.2 to +0.3 V. The $I_D$ vs $V_{SD}$ curves in Figure 2c reveal an enhancement of the transmembrane current for negative $V_G$. The current modulation is significant for both positive and negative $V_{SD}$, but with larger modulation for positive biases. Figure 2d plots $I_D$ as a function of $V_G$ for a series of $V_{SD}$ and reveals that at $V_{SD}$ = 0.5 V we observe a 10-fold modulation of the ionic current magnitude in the range of the applied gate voltages. To quantify the gating, we have also calculated the differential conductance ($G'=\Delta I_D/\Delta V_{SD}$) for the positive $V_{SD}$ values and confirmed the 10-fold enhancement of $I_D$ current as $V_G$ changes from -0.2 V to +0.3 V (Figure 2e).

The observation of the strong current enhancement for negative $V_G$ was observed in two additional independently fabricated stacks and nanopores (See Supporting Information Section 3). We verified that the effect of gating occurred in devices with symmetric thicknesses of the hBN layers as well as devices where the two hBN layers significantly differed in their thickness. These results confirmed that a graphite gate whose thickness does not exceed a few nm and, in some devices, constitutes only 1/10$^{th}$ of the entire stack thickness can effectively modulate the transmembrane current.

To understand the behavior of our ionic transistors, we considered a simple electrostatic model of the nanopore, with an asymmetry introduced by the negative hBN surface charge and the 3-electrode configuration. Based on the voltage gating observed in the experiments and cation selectivity of the nanopores, we analyzed in more detail the electrostatic interactions occurring in our system in different $V_{SD}$ and $V_G$ configurations (Figure 2 f-i) and proposed the mechanism of current modulation.

We begin with the analysis of the system when $V_{SD}$ and $V_G$ are zero (Figure 2f). In this case, ionic concentrations in the pore are entirely dominated by the presence of surface charges on the pore walls as in Figure 1e. The concentration of cations is enhanced in comparison to what one would expect if the pore was filled with bulk solution, which increases the pore conductance. The configuration in Figure 2f is our starting point to consider three cases with different arrangements of $V_{SD}$ and $V_G$: (i) $V_G$ < 0 and $V_{SD}$ > 0 (Figure 2g), (ii) $V_G$ < 0 and $V_{SD}$ < 0 (Figure 2h), and (iii) $V_G$ > 0 for both polarities of $V_{SD}$, (Figure 2i).

Figure 2g shows the schematic of electric potential and local cation concentration for $V_G$ < 0 and $V_{SD}$ > 0. Application of negative gate voltages will increase $K^+$ concentration near the gate beyond the enhancement that results from the presence of negative surface charges on hBN, hence leading to a higher conductance. It is important to note that in this three-electrode device, applying positive or negative $V_{SD}$ produces different electric potential (and ionic concentration) profiles in the pore due to the asymmetry of the source-to-gate, $V_{SG}$, and drain-to-gate, $V_{DG}$, potential differences. For example, when $V_G$ = -0.2 V and $V_{SD}$ = 0.5 V (Figure 2g), $V_{SG}$ is 0.7 V, and $V_{DG}$ is 0.2 V, both enhancing the cation concentration that already exists due the surface charge. On the other hand, when $V_{SD}$ < 0, e.g. $V_{SD}$ = -0.5 V, $V_{SG}$ = -0.3 V and the majority carriers (cations) are driven into the pore only from the drain side with $V_{DG}$ = 0.2 V, and exit through the source side. This results in a smaller relative increase in $K^+$ ion concentration and conductance when compared to $V_G$ = -0.2 V and $V_{SD}$ = 0.5 V. Due to the clear differences in the nanopore ionic concentrations for the same $V_G$ but opposite polarities of $V_{SD}$, and the stronger enhancement of $K^+$ concentration for positive $V_{SD}$, the current modulation will also be more significant for positive $V_{SD}$ values. Thus, the simple schemes of ionic concentrations for negative $V_G$ with positive (Figure 2g) or negative $V_{SD}$ (Figure 2h) provide an intuitive explanation for the asymmetry observed in the experimental $I_D$ vs $V_{SD}$ curves (Figure 2c).

Finally, a positive $V_G$ (Figure 2i) will repel $K^+$ ions away from the center of the device and create an ionic bipolar junction with two anti-parallel ionic diodes.[51, 52] For both polarities of $V_{SD}$, one of the diode junctions remains reverse biased, leading to an overall reduced concentration of ions in the nanopore, and, consequently, reduced transmembrane current. The $I_D$ vs $V_{SD}$ characteristics become more symmetric for more positive $V_G$, which points to the dominant role of the gate in ionic depletion. The experiments also reveal that when $V_G$ > 0, ionic concentrations become similar for both positive and negative $V_{SD}$ voltages, leading to more symmetric $I_D$ vs $V_{SD}$ curves.

The device shown in Figure 2 remained operational for ~2 months, which we attribute to the low gate voltages applied during the measurements. Our 3-electrode set-up also allowed us to quantify the leakage gate currents, whose detailed analysis is included in the Supporting Information file (Section 4). The measurements confirmed that the leakage current does not influence our transmembrane currents such that $V_G$ modulates ionic concentrations in the pore but does not contribute to $I_D$.

The salt concentration is a critical parameter that determines the magnitude of electrostatic modulation in nanofluidic systems.[1, 13, 16] In general, lower salt concentrations allow electrostatic interactions within the nanopore to dominate the conductance, whereas at high salt concentrations, the bulk properties of the solution control the transport behavior.[42] Therefore, it is important to determine the concentration range at which the gating effect is attainable in our devices. To accomplish this, we characterized the gating properties of device C (Figure 3a, b), by recording $I_D$ vs $V_{SD}$ curves for different gate voltages at KCl concentrations ranging from 1 mM to 1 M (See Supporting Information, Section 3). From these $I_D$-$V_{SD}$ curves, we determined the local differential conductance of the nanopore for positive $V_{SD}$ values (from 0.3 to 0.6 V) where the gating effects are the strongest (see Supporting Information, Figure S3) and plotted it as a function of the KCl concentration and $V_G$. Figure 3a shows that the gating effects are more significant at lower KCl concentrations. In addition, for all concentrations probed, the conductance increases with the magnitude of negative $V_G$ and decreases for more positive $V_G$ relative to the $V_G$ = 0 case. To further quantify the conductance gating with $V_G$, at each KCl concentration we calculated the modulation ratio $G'_{max}/G'_{min}$ (Figure 3b), defined as the conductance at $V_G$ = -0.3 V divided by that at $V_G$ = 0.3 V. The analysis in Figure 3b confirms that the current modulation becomes stronger as salt concentration decreases. It is noteworthy that the gating is already significant at 100 mM KCl, with a modulation ratio of 1.8 ± 0.1, and it continues to increase for lower concentrations reaching a maximum of 4.3 ± 0.4 at 1 mM KCl for this device. At 1 M KCl the modulation ratio of 1.3 ± 0.2 is the closest to 1 (no modulation).

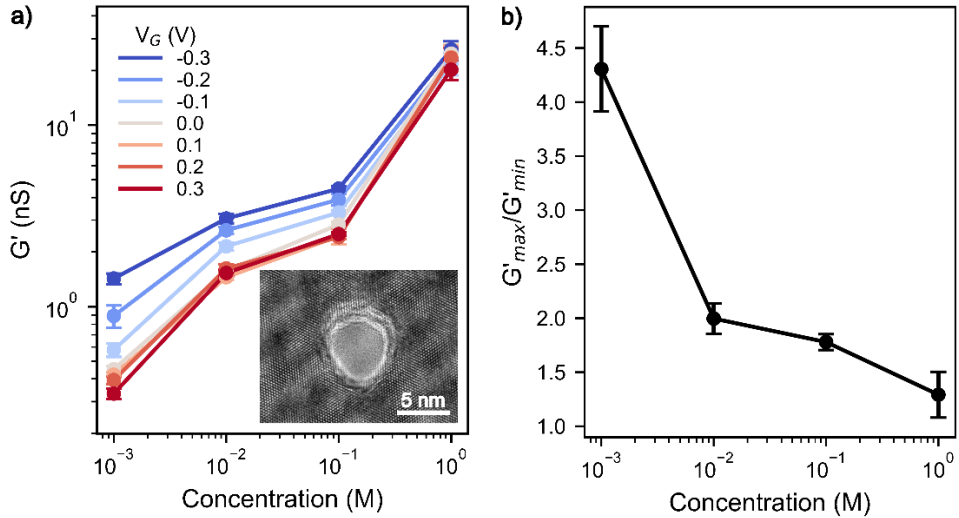

**Figure 3. Gating of ion current in vdW ionic transistors as a function of salt concentration.** *a) Local differential conductance as a function of salt concentration for Device C. The local conductance was calculated for the $V_{SD}$ range from 0.3 to 0.6V, where the modulation effects are the strongest. The inset shows TEM image of the nanopore. Device C consisted of two 11 nm thick hBN layers, and a 3.5 nm thick layer of graphene in the middle of the sandwich structure. b) Conductance ratio G' ($V_G$ = -0.3V)/ G'($V_G$ = 0.3V) for each KCl concentration. This ratio is used to quantify the dependence of the conductance modulation on the ionic concentration.*

**Continuum modeling of vdW ionic transistors reveals significant modulation of ionic concentrations with $V_G$.**

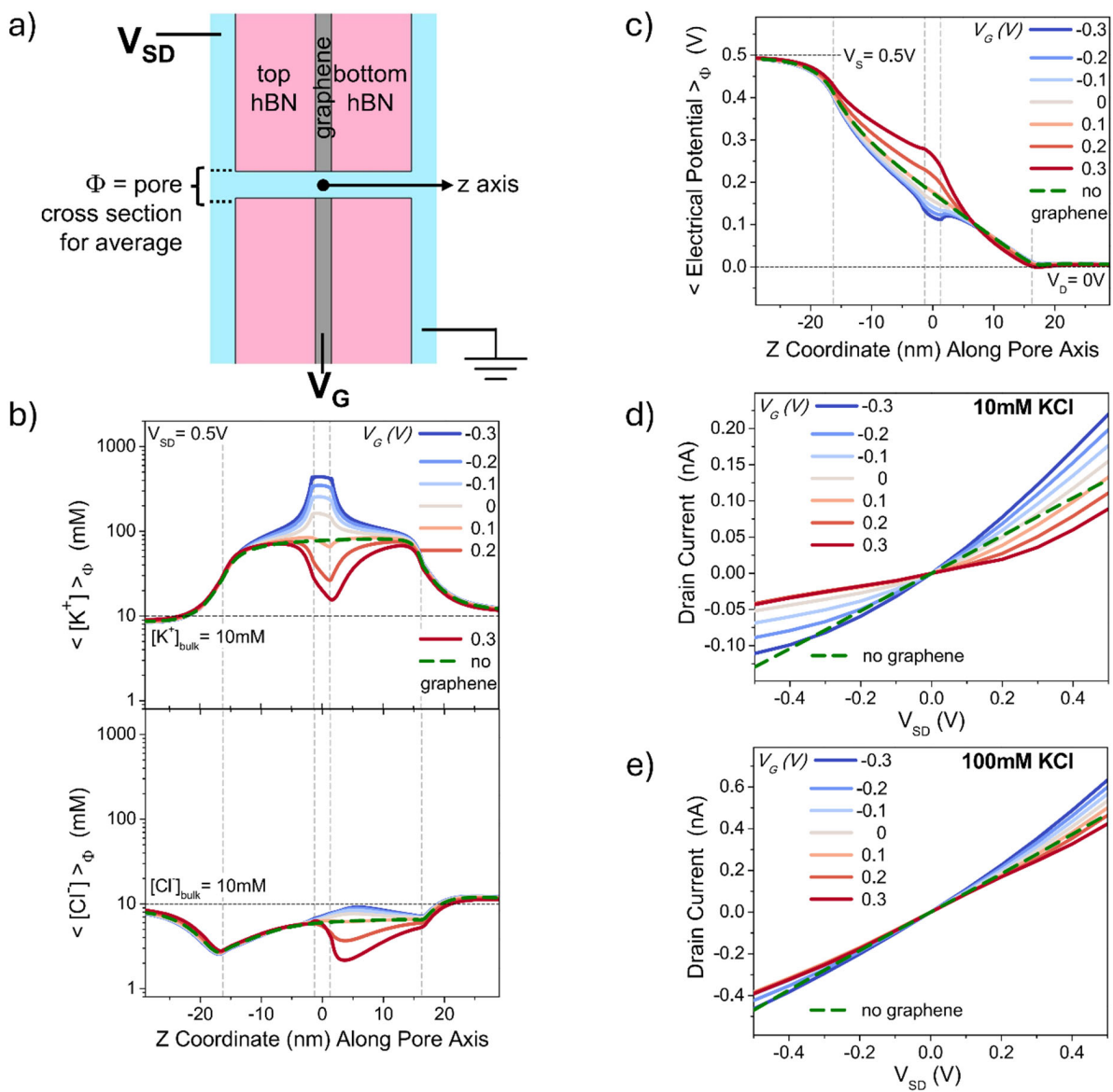

**Figure 4**. **COMSOL MultiPhysics finite element model of ionic concentrations and electric potential in a 5 nm in diameter vdW ionic transistor in 10 mM KCl as the bulk electrolyte**. *(a) The system consisted of two 15 nm thick layers of hBN and a 3 nm graphene layer. The surface charge density on all hBN surfaces was set at -10 mC/m$^2$. $V_{SD}$ is the voltage applied across the nanopore, and $V_G$ is the gate voltage at the graphene. (b) Profiles of ionic concentrations and (c) electric potential along the pore axis for different $V_G$ with $V_{SD}$ = 0.5V. The profiles in (b) and (c) were averaged over the pore cross section, as indicated in panel (a). (d, e) Current-voltage curves obtained from the model in 10 mM and 100 mM KCl, respectively. The modeling shown as a dashed green line (panels b-e) was obtained when the graphene region was changed to hBN and the applied gate voltage $V_G$ was disabled. All results were obtained with radially varying relative permittivity of the water inside the pore. The permittivity was set to a piecewise function of the radial coordinate (instead of the value of 80 used everywhere else for water) with magnitude of 2 for the 0.472 nm closest to the pore wall (the first interfacial water*

*layer), a value of 11 for the next 0.315 nm (the second interfacial layer), and a value 80 elsewhere in the middle of the pore (Figures S6, S7).*

To understand the gating effects in the vdW ionic transistor, we simulated the system using the Poisson-Nernst-Planck equations solved using the COMSOL Multiphysics package. Two 15 nm thick hBN layers were modeled considering physical properties of the material such as dielectric constant and a surface charge density of -10 mC/m$^2$ (Figure S8). Relative permittivity of the solution was assumed to be the lowest at the surface, in accordance with recent experimental and molecular dynamics simulations with hBN and other surfaces.[53, 54] Specifically, a piecewise relationship of the permittivity in the radial direction was assumed such that the magnitude of 2 was set at the 0.472 nm closest to the pore wall (corresponding to the first interfacial water layer), a value of 11 was used for the next 0.315 nm (the second interfacial layer), and a value 80 elsewhere in the middle of the pore (Figure S7).[53] A 3 nm graphene layer was declared as an electrode as shown in Figure 4a.

Figure 4b shows profiles of radially averaged ionic concentrations for $V_{SD}$ = 0.5 V and a range of $V_G$ between - 0.3 V and +0.3 V obtained from the COMSOL model. The model confirmed the predictions for gate voltage dependent ionic concentrations in the pore that were schematically analyzed in Figure 2. For negative $V_G$ and positive $V_{SD}$, concentration of K$^+$ ions is enhanced in the whole volume of the pore, reaching the highest values at the gate region, while the concentration of Cl$^-$ is depleted in accordance with the expected cation selectivity of the system.[41, 55-57]

As predicted by the schemes in Figure 2, the voltage configuration with $V_{SD}$ = -0.5 V and $V_G$ = -0.3 V leads to a weaker enhancement of K$^+$ ions concentration inside the pore compared to $V_{SD}$ = 0.5 V, and therefore, weaker current modulation (see Figure 4b and Supporting Information, Figures S9, S10). Interestingly, even though potassium concentration is enhanced above the bulk value at $V_G$ < 0 for both $V_{SD}$ = -0.5 V and $V_{SD}$ = 0.5 V cases, at $V_{SD}$ = -0.5 V potassium ions assume the lowest concentrations in the gate region, even lower than in the configuration without any gate ("no graphene"), Figure S9. This counterintuitive finding can be understood through the effect of concentration polarization[41, 57-59] due to the ion selectivity of the gate and hBN regions. This leads to the formation of a depletion zone not only at the drain side (right-hand side in Figure 4a, where K$^+$ ions enter the pore in this case) but also at the gate, since K$^+$ ions are then sourced from the gate region to pass towards the source. This conclusion is also supported by the profile of electric potential in Figure S9b for $V_{SD}$ = -0.5 V that reveals formation of a barrier at the gate, even at negative gate voltages.

The modeling also provided a direct insight into what happens when a positive gate is applied. In accordance with our intuition based on a similar system of a bipolar junction,[51, 52] $V_G$ > 0 forms a depletion zone whose extent is voltage dependent and limits the transmembrane current. For $V_{SD}$ = 0.5 V and $V_G$ = 0.3 V, concentration of potassium ions remains above the bulk value in the whole pore volume but is the lowest in a zone that extends from the axial position of ~-5 nm to +~10 nm (Figure 4). For the configuration of $V_{SD}$ = -0.5 V and $V_G$ = +0.3 V, on the other hand, due to the larger potential difference between source and gate, $V_{SG}$ = 0.8 V, potassium ions are depleted below the bulk concentration in a gate region, at the axial positions between -10 nm and 5 nm (Figure S9b for 10 mM KCl). These results underscore the intrinsic asymmetry of the 3-electrode system.

The distribution of electric potential along the pore axis for a range of gate voltages is shown in Figure 4c. The results revealed that $V_G$ significantly influenced the local electric potential in the pore leading to the formation of a well (barrier) for potassium ions at $V_G < 0$ ($V_G > 0$) for $V_{SD} = 0.5$ V. The presence of the graphene layer is responsible for the broken electrochemical symmetry of the system observed as rectification of ion current and gating that depends on the polarity of $V_{SD}$. The calculated current-voltage curves in Figure 4 d,e qualitatively agree with the experimental data and predict enhancement (decrease) of the current for negative (positive) gate voltages.

It is important to note that the quantitative agreement of the continuum model with experimental data was significantly improved by introducing the diminished relative permittivity of the solution next to the pore walls. Figure S11 shows results from the same model but with solution permittivity set to 80 in the whole pore volume. When homogeneous, bulk permittivity was assumed, the modulation of ion concentrations reached unphysically high values and the gating at positive $V_G$ predicted a nearly complete switching off the currents that was not observed in experiments.

**Molecular dynamics simulations support reduced relative permittivity of water in the ionic transistors and the interdependence of water and ionic distributions**

To reinforce COMSOL assumptions and predictions, and to gain additional molecular-level insight into the gating mechanism, we performed classical molecular dynamics (MD) simulations using empirical force fields on a smaller, representative model of the vdW ionic transistor. Figure 5a shows snapshots of the simulation setup, which included prescribed surface charge densities on the hBN layers, a source–drain electric field equivalent to +0.5 V applied across the pore, and partial charges applied to the graphene sheets to impose positive and negative gate potentials. The model consisted of a 1 nm diameter nanopore drilled through an hBN-graphene-hBN stack connecting two water reservoirs containing 100 mM KCl with a 75 Å long nanopore (Figure 5a). Further details of the simulations can be found in the Methods section.

Figure 5b shows how water orientation within the pore responds to the applied gate potential through the combined effects of surface charging, the source-drain electric field, and confinement. Water dipole orientations are defined by the angle between the molecular bisector, weighted toward the hydrogen atoms, and the source–drain field along the positive z direction, such that alignment toward a negatively charged surface corresponds to hydrogens pointing toward the interface. As shown in Figure S12, in the absence of surface charges water molecules preferentially orient radially within the nanopore. Introducing partial charges on the hBN sheets, while leaving the graphene uncharged and applying no source–drain field, alters the preferred water orientation to align with the nanopore axis throughout the pore, except in the graphene region where no surface charge is present. Under a negative gate potential, water dipoles align parallel to the pore axis below the pore midpoint and antiparallel above it, revealing a clear spatial asymmetry in orientation. This behavior arises from the superposition of the axial source–drain field and the local electrostatic fields generated by the negatively charged graphene surfaces, which reverse sign across the pore midpoint and drive a corresponding flip in the preferred water dipole orientation. In contrast, under a positive gate field (positive partial charges on the graphene sheets), water in both the hBN and graphene regions predominantly aligns parallel to the source–drain field, reflecting the dominance of the axial driving field over surface electrostatics. Here, perpendicular alignment is confined to the first hBN–graphene interface, where strong local fields and interfacial polarization locally outweigh the axial field.

Overall, partial surface charges on the nanosheets drive water molecules to align along the pore axis, suppressing radial dipole fluctuations and thereby reducing the out-of-plane dielectric response.[46] This microscopic behavior supports the reduced relative permittivity assigned to nanoscale confined water along the confinement direction, as assumed in the COMSOL simulations. The observed water ordering is a local interfacial effect and is therefore expected to persist near the surface for larger pore widths. However, increased electrostatic screening in wider pores would reduce the spatial extent of this ordering and its impact on the pore interior.

Figure 5c presents the ion ($K^+$, $Cl^-$) distributions within the pore as a function of position for positive and negative surface charging applied to the graphene sheet. Here, the reported frequency reflects both the number of ions and their residence time within each spatial bin. Under a negative graphene surface charge, $K^+$ ions preferentially accumulated near the graphene region and in the adjacent hBN region, whereas $Cl^-$ ions were effectively excluded from the pore, consistent with the strongly negative electrostatic environment arising from negatively charged graphene and hBN surfaces. In contrast, applying a positive graphene surface charge substantially altered the ionic organization: $K^+$ ions remained the dominant species but shifted toward the hBN region, and Cl- ions entered the pore and preferentially accumulated near the graphene/hBN interface. Notably, this accumulation coincides with the region where water molecules exhibit a perpendicular orientational ordering, suggesting a possible strong coupling between interfacial water structure and ion localization. The ion distributions agree well with the COMSOL concentration profiles for the 100 mM KCl model, indicating that the COMSOL simulations capture the relevant microscopic physics. Ultimately, the MD simulations demonstrate that surface charges in the hBN–graphene–hBN stack strongly couple water structure and dynamics to ion localization, underscoring the central role of water-mediated effects in governing ion transport through nanopores.

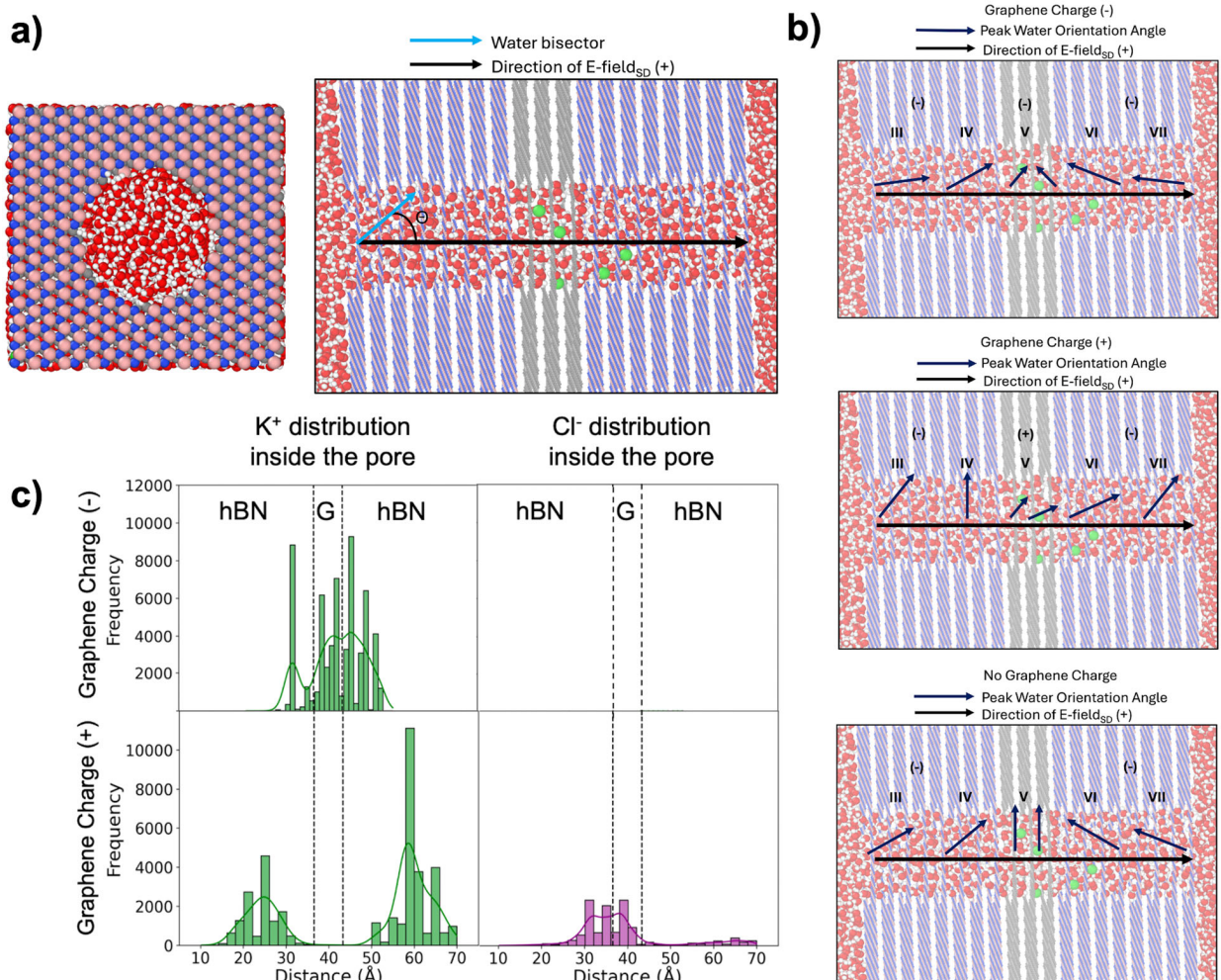

**Figure 5. Classical molecular dynamics simulations of a 1 nm-diameter nanopore drilled through an hBN-graphene-hBN stack in 100 mM KCl electrolyte**. It is a smaller, representative model of the experimental system. *(a) Top-down snapshot of system setup and snapshot of the nanopore drilled through the hBN-graphene-hBN stack, illustrating the applied source–drain electric field and the water bisector vector used to define molecular orientation. (b) Pictorial representation of the peak water orientation angles along the nanopore, resolved by spatial region, with applied surface charges indicated for a positive, negative, and no surface charge on the graphene sheets; peak positions are extracted from the orientation distributions shown in Figure S12. (c) Ion frequency distributions along the nanopore over 200 ns for simulations with positive and negative graphene surface charges.*

**Conclusions**

In summary, we demonstrate ionic transistors based on van der Waals hBN-graphene-hBN stacks that deliever strong, reversible modulation of the ionic current at low voltages, and at salt concentrations up to 100 mM KCl. Modulations of 10 fold and 2 fold were measured at 10 mM and 100 mM KCl, respectively at $V_G$ as low as 0.3 V. The three-electrode system exhibits an instrinscilly broken symmetry of the electrochemcical potential where the effects of gate voltage, $V_G$, depend on the polarity and magnitude of the source-drain voltage. This pronounced asymmetry is consistent with a negatively-charged, high aspect ratio hBN pore combined with an internal few-nanometer gate whose polarity sets the local ion density: negative $V_G$ increases $K^+$ concentration and enhances the current, while positive $V_G$ induces local ionic depletion and suppresses the transmembrane conductance. This mechanism is further elucidated by a continuous model based on the coupled Poisson–Nernst–Planck equations that reproduces the asymetric $I_D$-$V_{SD}$ curves and reveals how gate polarity sculpts local ion profiles. Molecular dynamics simulations demonstrate that water-mediated effects play a central role in governing ion transport through nanopores and support a reduced effective permittivity for the confined water. These insights translate into practical design rules linking pore geometry, surface charge, salt concentration, water mediated effects, and gate bias, enabling the next generation of low-power ionic circuits that operate at physiological ionic strength.

**Acknowledgments**
We acknowledge support from the Center for Enhanced Nanofluidic Transport (CENT), an Energy Frontier Research Center funded by DOE, Office of Science, Basic Energy Sciences under Award DE-SC0019112 (A.B-A., M.S., M.B., F.A, T. A.P., J.S-Y, Z.S.S). Development of the Comsol model was supported by the National Science Foundation CHE, MPS 2200524 (E.C., D.V., Z.S.S). The authors acknowledge the use of facilities and instrumentation at the UC Irvine Materials Research Institute (IMRI), which is supported in part by the National Science Foundation through the UC Irvine Materials Research Science and Engineering Center (DMR-2011967) as well as the use of facilities and instrumentation at the Integrated Nanosystems Research Facility (INRF) and RapidTech at the UC Irvine Samueli School of Engineering. Computing support was provided by the Lawrence Livermore National Laboratory Institutional Computing Grand Challenge program. Work at LLNL was performed under the auspices of the U.S. Department of Energy by Lawrence Livermore National Laboratory under Contract DE-AC52-07NA27344. We are grateful to Prof. Charles R. Martin for helpful discussions.

**Supporting Information**
The Supporting Information is available free of charge at: https://pubs.acs.org
Additional experimental and modeling of ionic transistors based on heterogeneous stacks of vdW materials, and details of the experimental set-up.

**Methods**

**Device assembly**
The devices were fabricated on $SiN_x$/Si TEM grids with a 50 x 50 µm large and 200 nm thick $SiN_x$ membrane and a single 5 µm diameter pore (Norcada, Inc.). All the graphene and hBN layers were mechanically exfoliated from bulk crystals. The stacks were fabricated by the dry transfer method[27] using stamps made of a polycarbonate (PC) film on top of a polydimethylsiloxane (PDMS) dome on a glass slide. Electron Beam Lithography (EBL) was used to write the patterns of the fine contacts using a layer of poly(methyl methacrylate) (PMMA) resist. For this purpose, PMMA 950 A5 was spun for 2 minutes at 2800 rpm. The EBL patterns were written at 1.6 nA with 30 kV excitation and then developed for 3 to 4 minutes in a cold mixture of 3:1 isopropyl alcohol (IPA)/water. After writing and developing the patterns, reactive ion etching (RIE) was used to expose the graphene with the following parameters: a flow of 10 standard cubic centimeters per minute (SCCM) of $SF_6$, 2 SCCM of $O_2$ and 30W of radio frequency power, at 100 mTorr for 30 s.[60] As the next step, 2 nm of Cr and 100 nm of Au were deposited in an electron beam evaporator system at the rate of 1 Å/s. Liftoff was performed by soaking the sample in acetone for 1 to 2 hours and agitating with a pipette. For the large contacts on the other hand, a pre-patterned polyamide shadow mask was placed on the sample and a second Cr and Au evaporation was carried to deposit the contacts. To pasivate the metallic contacts, a 170 nm layer of $SiO_2$ was deposited by PECVD at 250 °C. A comercial glass etcher was subsequently used to expose the outter ends of the metallic contacts. After that, a window opening was patterned on PMMA using EBL in the micropore zone of the sample, followed by wet etching a window in the $SiO_2$ using BOE 6:1 for a few minutes, exposing the stack only in the etched region. Finally, a single nanopore was drilled using a JEOL 2100f Transmission Electron Microscope with an accelerating voltage of 200 kV and 0.5 nA probe current.[40, 41] Drilling occurred by narrowing the beam and dwelling it on a single point on the surface for ~2 min. After drilling, the opening diameters of the nanopores were measured from TEM micrographs.

**Device Electrochemical Measurement**
The devices were measured at room temperature in a PDMS electrolytic cell adapted to incorporate a custom flexible polyimide PCB board that allowed us to establish electrical connection with the devices (Supporting Information, Section 1). Two small carbon tape squares cut with a microknife (Fine Science Tools) were placed on the PCB center contacts to ensure strong electrical connection to the chip contacts. Two 50 μm thick PDMS (Nacalai USA, Inc.) laser cut (Epilog Fusion M2 Laser Cutter) rectangular o-rings with center opening of 200 μm diameter were placed over the center opening of the PCB to ensure good fluidic seal between chip and the electrolytic cell channel. The current into the drain was measured with a Keithley 6487 picoammeter that simultaneously sourced the source-drain voltage. The source and drain were the fluidic reservoirs of the electrolytic cell, and two Ag/AgCl electrodes were used for the electrochemical measurements. The gate voltage was applied with a Keithley 2450 SMU that also measured the gate current using SMA connections with the PCB. The grounds of these two Keithleys were connected. All devices were first probed in a two-electrode set-up with source-drain voltage, and the gate voltage floating. We began the measurements in 1 M KCl with 50% ethanol/50% deionized water (Milli-Q IQ 7000), because high concentrated solutions and the presence of ethanol facilitate wetting of nanopores.[61] Wetting was determined by comparing the experimentally measured conductance with expected conductance based on the pore diameter as measured by TEM as well as stability of ion current signal. Once the measurements in 1 M KCl 50% ethanol/50% water were successful, recordings with 100% water and lower KCl concentrations were performed in 2 and 3-electrode configurations. Current-voltage curves were recorded using custom written Labview programs. At least three voltage scans were recorded and the data shown represent arithmetic average of recordings at each voltage. pH of KCl solutions was adjusted with 0.1 M HCl or NaOH.

**Comsol simulations**
Models were developed to investigate ionic transport through single nanopores with an inner gating region. Two-dimensional axisymmetric geometries were constructed and solved in COMSOL Multiphysics using the coupled Poisson–Nernst–Planck equations.[41] Including the Navier–Stokes equations had a negligible impact on the results, and the effects of fluid flow were therefore omitted. Details of the model are included in Supporting Information.

**Molecular Dynamics Simulation**

Classical MD simulations were carried out with the LAMMPS simulation package.[62] The system was solvated using the SPC/E water model, with O-H bonds constrained using the SHAKE algorithm. Force-field parameters for hBN were taken from Won and Aluru,[63] while graphene and ion interactions were modeled using the OPLS-AA force field.[64] The model consisted of a 1 nm diameter nanopore drilled through an hBN-graphene-hBN stack connecting two water reservoirs containing 100 mM KCl. The system's dimension is 37.6 Å x 39.6 Å x197.35 Å. The stack comprises eight hBN layers on each side of a three-layer graphene stack. A source-drain electric field equivalent to +0.5 V was applied in the Z direction defined to the region of the nanopore (with a length of 75 Å) and additional gate fields of +/- 0.3 V were applied to the graphene sheets. To impose positive and negative gate fields, partial charges of +0.01e and -0.01e, respectively, were assigned to each atom in the graphene sheet, following prior work.[61] In addition, an extra -0.01e charge per atom was applied to the hBN layer to reproduce a negatively charged hBN surface, consistent with experimental observations and COMSOL modeling. Long-range electrostatics were treated using the particle-particle-particle-mesh (PPPM) method.[64] After energy minimization, the system was equilibrated in the NPT ensemble using a Berendsen barostat.[65] Production simulations of 200 ns were then performed in the NVT ensemble with a Nose-Hoover thermostat,[66] maintaining the temperature at 298.15 K.

**References**
(1) Schoch, R. B.; Han, J. Y.; Renaud, P. Transport Phenomena in Nanofluidics. *Rev. Mod. Phys.* **2008**, *80* (3), 839-883. DOI: 10.1103/RevModPhys.80.839.
(2) Aluru, N. R.; Aydin, F.; Bazant, M. Z.; Blankschtein, D.; Brozena, A. H.; de Souza, J. P.; Elimelech, M.; Faucher, S.; Fourkas, J. T.; Koman, V. B.; et al. Fluids and Electrolytes under Confinement in Single-Digit Nanopores. *Chem. Rev.* **2023**, *123* (6), 2737-2831. DOI: 10.1021/acs.chemrev.2c00155.
(3) Faucher, S.; Aluru, N.; Bazant, M. Z.; Blankschtein, D.; Brozena, A. H.; Cumings, J.; Pedro de Souza, J.; Elimelech, M.; Epsztein, R.; Fourkas, J. T.; et al. Critical Knowledge Gaps in Mass Transport through Single-Digit Nanopores: A Review and Perspective. *J. Phys. Chem. C* **2019**, *123* (35), 21309-21326. DOI: 10.1021/acs.jpcc.9b02178.
(4) Siwy, Z. S.; Howorka, S. Engineered Voltage-Responsive Nanopores. *Chem. Soc. Rev.* **2010**, *39*, 1115-1132.
(5) Daiguji, H.; Oka, Y.; Shirono, K. Nanofluidic Diode and Bipolar Transistor. *Nano Lett.* **2005**, *5* (11), 2274-2280. DOI: 10.1021/nl051646y.
(6) Zhang, Z.; Huang, X.; Qian, Y.; Chen, W.; Wen, L.; Jiang, L. Engineering Smart Nanofluidic Systems for Artificial Ion Channels and Ion Pumps: From Single-Pore to Multichannel Membranes. *Adv. Mater.* **2020**, *32* (4), 1904351. DOI: https://doi.org/10.1002/adma.201904351.
(7) Hille, B. Ion Channels of Excitable Membranes. **2001**, Sinauer Associates: Sunderland.
(8) Ritchie, G. J. *Transistor Circuit Techniques: Discrete and Integrated*; Van Nostrand Reinhold (UK), 1983.
(9) Sze, S. M.; Li, Y.; Ng, K. K. *Physics of Semiconductor Devices*; Wiley, 2021.

(10) Mei, T.; Liu, W.; Xu, G.; Chen, Y.; Wu, M.; Wang, L.; Xiao, K. Ionic Transistors. *ACS Nano* **2024**, *18* (6), 4624-4650. DOI: 10.1021/acsnano.3c06190.
(11) Tybrandt, K.; Forchheimer, R.; Berggren, M. Logic Gates Based on Ion Transistors. *Nat Commun.* **2012**, *3* (1), 871. DOI: 10.1038/ncomms1869.
(12) Lucas, R. A.; Lin, C.-Y.; Baker, L. A.; Siwy, Z. S. Ionic Amplifying Circuits Inspired by Electronics and Biology. *Nat. Commun.* **2020**, *11* (1), 1568. DOI: 10.1038/s41467-020-15398-3.
(13) Nam, S. W.; Rooks, M. J.; Kim, K. B.; Rossnagel, S. M. Ionic Field Effect Transistors with Sub-10 nm Multiple Nanopores. *Nano Lett.* **2009**, *9*, 2044-2048.
(14) Guan, W.; Fan, R.; Reed, M. A. Field-Effect Reconfigurable Nanofluidic Ionic Diode. *Nat. Commun.* **2011**, *2*, 506.
(15) Karnik, R.; Fan, R.; Yue, M.; Li, D.; Yang, P.; Majumdar, A. Electrostatic Control of Ions and Molecules in Nanofluidic Transistors. *Nano Lett.* **2005**, *5* (5), 943-948. DOI: 10.1021/nl050493b.
(16) Liu, W.; Mei, T.; Cao, Z.; Li, C.; Wu, Y.; Wang, L.; Xu, G.; Chen, Y.; Zhou, Y.; Wang, S.; et al. Bioinspired Carbon Nanotube–Based Nanofluidic Ionic Transistor with Ultrahigh Switching Capabilities for Logic Circuits. *Sci. Adv.* **2024**, *10* (11), eadj7867. DOI: doi:10.1126/sciadv.adj7867.
(17) Xue, Y.; Xia, Y.; Yang, S.; Alsaid, Y.; Fong, K. Y.; Wang, Y.; Zhang, X. Atomic-Scale Ion Transistor with Ultrahigh Diffusivity. *Science* **2021**, *372* (6541), 501-503. DOI: doi:10.1126/science.abb5144.
(18) Tsutsui, M.; Hsu, W.-L.; Garoli, D.; Leong, I. W.; Yokota, K.; Daiguji, H.; Kawai, T. Gate-All-Around Nanopore Osmotic Power Generators. *ACS Nano* **2024**, *18* (23), 15046-15054. DOI: 10.1021/acsnano.4c01989.
(19) Sun, G.; Senapati, S.; Chang, H.-C. High-Flux Ionic Diodes, Ionic Transistors and Ionic Amplifiers Based on External Ion Concentration Polarization by an Ion Exchange Membrane: A New Scalable Ionic Circuit Platform. *Lab Chip* **2016**, *16* (7), 1171-1177, 10.1039/C6LC00026F. DOI: 10.1039/C6LC00026F.
(20) Qiu, H.; Zhou, W.; Guo, W. Nanopores in Graphene and Other 2D Materials: A Decade's Journey toward Sequencing. *ACS Nano* **2021**, *15* (12), 18848-18864. DOI: 10.1021/acsnano.1c07960.
(21) Macha, M.; Marion, S.; Nandigana, V. V. R.; Radenovic, A. 2D Materials as an Emerging Platform for Nanopore-Based Power Generation. *Nat. Rev. Mater.* **2019**, *4* (9), 588-605. DOI: 10.1038/s41578-019-0126-z.
(22) Xing, X.-L.; Li, W.; Guo, L.-R.; Wang, K.; Ma, Y.-z.; Zhao, Q.; Ji, L. Nanopores in 2D Materials and Their Applications in Single Molecule Analysis. *TrAC Trends in Analytical Chemistry* **2024**, *179*, 117863. DOI: https://doi.org/10.1016/j.trac.2024.117863.
(23) Wang, L.; Boutilier, M. S. H.; Kidambi, P. R.; Jang, D.; Hadjiconstantinou, N. G.; Karnik, R. Fundamental Transport Mechanisms, Fabrication and Potential Applications of Nanoporous Atomically Thin Membranes. *Nat. Nanotechnol.* **2017**, *12* (6), 509-522. DOI: 10.1038/nnano.2017.72.
(24) Sahu, S.; Zwolak, M. Colloquium: Ionic Phenomena in Nanoscale Pores Through 2D Materials. *Rev. Mod. Phys.* **2019**, *91* (2), 021004. DOI: 10.1103/RevModPhys.91.021004.
(25) Chen, S.; Huang, S.; Son, J.; Han, E.; Watanabe, K.; Taniguchi, T.; Huang, P. Y.; King, W. P.; van der Zande, A. M.; Bashir, R. Detecting DNA Translocation Through a Nanopore Using a van der Waals heterojunction diode. *Proc. Natl. Acad. Sci. U.S.A.* **2025**, *122* (18), e2422135122. DOI: doi:10.1073/pnas.2422135122.
(26) Li, X.; Lin, S.; Lin, X.; Xu, Z.; Wang, P.; Zhang, S.; Zhong, H.; Xu, W.; Wu, Z.; Fang, W. Graphene/h-BN/GaAs Sandwich Diode as Solar Cell and Photodetector. *Opt. Express* **2016**, *24* (1), 134-145. DOI: 10.1364/OE.24.000134.
(27) Purdie, D. G.; Pugno, N. M.; Taniguchi, T.; Watanabe, K.; Ferrari, A. C.; Lombardo, A. Cleaning Interfaces in Layered Materials Heterostructures. *Nat. Commun.* **2018**, *9* (1), 5387. DOI: 10.1038/s41467-018-07558-3.

(28) Yuan, Z.; Liang, Z.; Yang, L.; Zhou, D.; He, Z.; Yang, J.; Wang, C.; Jiang, L.; Guo, W. Light-Driven Ionic and Molecular Transport through Atomically Thin Single Nanopores in MoS2/WS2 Heterobilayers. *ACS Nano* **2024**, *18* (35), 24581-24590. DOI: 10.1021/acsnano.4c09555.
(29) Schwestka, J.; Inani, H.; Tripathi, M.; Niggas, A.; McEvoy, N.; Libisch, F.; Aumayr, F.; Kotakoski, J.; Wilhelm, R. A. Atomic-Scale Carving of Nanopores into a van der Waals Heterostructure with Slow Highly Charged Ions. *ACS Nano* **2020**, *14* (8), 10536-10543. DOI: 10.1021/acsnano.0c04476.
(30) Geim, A. K.; Grigorieva, I. V. Van der Waals heterostructures. *Nature* **2013**, *499* (7459), 419-425. DOI: 10.1038/nature12385.
(31) Wu, X.; Yang, R.; Chen, X.; Liu, W. Fabrication of Nanopore in MoS$_2$-Graphene vdW Heterostructure by Ion Beam Irradiation and the Mechanical Performance. *Nanomaterials (Basel)* **2022**, *12* (2). DOI: 10.3390/nano12020196 From NLM.
(32) Liang, S.; An, X.; Liu, T.; Li, Y.; Gao, F.; Yin, Z.; Xia, L.; Zhang, J. Atomically Suspended Graphene/MoS$_2$ Heterojunction Driven High-Resolution Air Pressure Sensor for Fall Detection. *Nano Lett.* **2026**, *26* (3), 936-943. DOI: 10.1021/acs.nanolett.5c04403.
(33) Ma, L.; Liu, Z.; Man, J.; Li, J.; Siwy, Z. S.; Qiu, Y. Modulation Mechanism of Ionic Transport Through Short Nanopores by Charged Exterior Surfaces. *Nanoscale* **2023**, *15* (46), 18696-18706. 10.1039/D3NR04467J.
(34) Graf, M.; Lihter, M.; Altus, D.; Marion, S.; Radenovic, A. Transverse Detection of DNA Using a MoS2 Nanopore. *Nano Lett.* **2019**, *19* (12), 9075-9083. DOI: 10.1021/acs.nanolett.9b04180.
(35) Feng, J. D.; Liu, K.; Bulushev, R. D.; Khlybov, S.; Dumcenco, D.; Kis, A.; Radenovic, A. Identification of Single Nucleotides in MoS$_2$ Nanopores. *Nat. Nanotechnol.* **2015**, *10* (12), 1070-1076.
(36) Schneider, G. F.; Kowalczyk, S. W.; Calado, V. E.; Pandraud, G.; Zandbergen, H. W.; Vandersypen, L. M. K.; Dekker, C. DNA Translocation through Graphene Nanopores. *Nano Lett.* **2010**, *10* (8), 3163-3167. DOI: 10.1021/nl102069z.
(37) Merchant, C. A.; Healy, K.; Wanunu, M.; Ray, V.; Peterman, N.; Bartel, J.; Fischbein, M. D.; Venta, K.; Luo, Z. T.; Johnson, A. T. C.; et al. DNA Translocation through Graphene Nanopores. *Nano Lett.* **2010**, *10* (8), 2915-2921. DOI: 10.1021/nl101046t.
(38) Chou, Y.-C.; Lin, C.-Y.; Castan, A.; Chen, J.; Keneipp, R.; Yasini, P.; Monos, D.; Drndić, M. Coupled Nanopores for Single-Molecule Detection. *Nat. Nanotechnol.* **2024**, *19* (11), 1686-1692. DOI: 10.1038/s41565-024-01746-7.
(39) Smeets, R. M. M.; Keyser, U. F.; Krapf, D.; Wu, M. Y.; Dekker, N. H.; Dekker, C. Salt Dependence of Ion Transport and DNA Translocation Through Solid-State Nanopores. *Nano Lett.* **2006**, *6* (1), 89-95. DOI: 10.1021/nl052107w.
(40) Storm, A. J.; Chen, J. H.; Ling, X. S.; Zandbergen, H. W.; Dekker, C. Fabrication of Solid-State Nanopores with Single-Nanometre Precision. *Nat. Mater.* **2003**, *2* (8), 537-540. DOI: 10.1038/nmat941.
(41) Cao, E.; Cain, D.; Silva, S.; Siwy, Z. S. Ion Concentration Polarization Tunes Interpore Interactions and Transport Properties of Nanopore Arrays. *Adv. Funct. Mater.* **2024**, *34* (11), 2312646. DOI: https://doi.org/10.1002/adfm.202312646.
(42) Stein, D.; Kruithof, M.; Dekker, C. Surface-Charge-Governed Ion Transport in Nanofluidic Channels. *Phys. Rev. Lett.* **2004**, *93* (3), 035901. DOI: 10.1103/PhysRevLett.93.035901.
(43) Ho, C.; Qiao, R.; Heng, J. B.; Chatterjee, A.; Timp, R. J.; Aluru, N. R.; Timp, G. Electrolytic Transport Through a Synthetic Nanometer-Diameter Pore. *Proc. Nat. Acad. Sci. U.S.A.* **2005**, *102* (30), 10445-10450. DOI: doi:10.1073/pnas.0500796102.
(44) van den Hout, M.; Hall, A. R.; Wu, M. Y.; Zandbergen, H. W.; Dekker, C.; Dekker, N. H. Controlling Nanopore Size, Shape and Stability. *Nanotechnology* **2010**, *21* (11), 115304. DOI: 10.1088/0957-4484/21/11/115304.

(45) Lee, C.; Joly, L.; Siria, A.; Biance, A.-L.; Fulcrand, R.; Bocquet, L. Large Apparent Electric Size of Solid-State Nanopores Due to Spatially Extended Surface Conduction. *Nano Lett.* **2012**, *12* (8), 4037-4044. DOI: 10.1021/nl301412b.
(46) Siria, A.; Poncharal, P.; Biance, A. L.; Fulcrand, R.; Blase, X.; Purcell, S. T.; Bocquet, L. Giant Osmotic Energy Conversion Measured in a Single Transmembrane Boron Nitride Nanotube. *Nature* **2013**, *494* (7438), 455-458. DOI: 10.1038/nature11876.
(47) Weber, M.; Koonkaew, B.; Balme, S.; Utke, I.; Picaud, F.; Iatsunskyi, I.; Coy, E.; Miele, P.; Bechelany, M. Boron Nitride Nanoporous Membranes with High Surface Charge by Atomic Layer Deposition. *ACS Appl. Mater. Interfaces* **2017**, *9* (19), 16669-16678. DOI: 10.1021/acsami.7b02883.
(48) Wang, Y.; Luo, H.; Advincula, X. R.; Zhao, Z.; Esfandiar, A.; Wu, D.; Fong, K. D.; Gao, L.; Hazrah, A. S.; Taniguchi, T.; et al. Spontaneous Surface Charging and Janus Nature of the Hexagonal Boron Nitride–Water Interface. *J. Am. Chem. Soc.* **2025**, *147* (33), 30107-30116. DOI: 10.1021/jacs.5c07827.
(49) Yazda, K.; Bleau, K.; Zhang, Y.; Capaldi, X.; St-Denis, T.; Grutter, P.; Reisner, W. W. High Osmotic Power Generation via Nanopore Arrays in Hybrid Hexagonal Boron Nitride/Silicon Nitride Membranes. *Nano Lett.* **2021**, *21* (10), 4152-4159. DOI: 10.1021/acs.nanolett.0c04704.
(50) Cantley, L.; Swett, J. L.; Lloyd, D.; Cullen, D. A.; Zhou, K.; Bedworth, P. V.; Heise, S.; Rondinone, A. J.; Xu, Z.; Sinton, S.; et al. Voltage Gated Inter-Cation Selective Ion Channels From Graphene Nanopores. *Nanoscale* **2019**, *11* (20), 9856-9861. DOI: 10.1039/C8NR10360G.
(51) Daiguji, H.; Yang, P.; Majumdar, A. Ion Transport in Nanofluidic Channels. *Nano Lett.* **2004**, *4* (1), 137-142. DOI: 10.1021/nl0348185.
(52) Kalman, E.; Vlassiouk, I.; Siwy, Z. S. Nanofluidic Bipolar Transistors. *Adv. Mater.* **2008**, *20*, 293-297.
(53) Fumagalli, L.; Esfandiar, A.; Fabregas, R.; Hu, S.; Ares, P.; Janardanan, A.; Yang, Q.; Radha, B.; Taniguchi, T.; Watanabe, K.; et al. Anomalously Low Dielectric Constant of Confined Water. *Science* **2018**, *360* (6395), 1339-1342. DOI: doi:10.1126/science.aat4191.
(54) Dufils, T.; Schran, C.; Chen, J.; Geim, A. K.; Fumagalli, L.; Michaelides, A. Origin of Dielectric Polarization Suppression in Confined Water from First Principles. *Chem. Sci.* **2024**, *15* (2), 516-527. DOI: 10.1039/D3SC04740G.
(55) Lucas, R. A.; Siwy, Z. S. Tunable Nanopore Arrays as the Basis for Ionic Circuits. *ACS Appl. Mater. Interfaces* **2020**, *12* (50), 56622-56631. DOI: 10.1021/acsami.0c18574.
(56) Li, M.; Anand, R. K. Recent Advancements in Ion Concentration Polarization. *Analyst* **2016**, *141* (12), 3496-3510. DOI: 10.1039/C6AN00194G.
(57) Zangle, T. A.; Mani, A.; Santiago, J. G. Theory and Experiments of Concentration Polarization and Ion Focusing at Microchannel and Nanochannel Interfaces. *Chem. Soc. Rev.* **2010**, *39* (3), 1014-1035. DOI: 10.1039/B902074H.
(58) Freger, V. Selectivity and Polarization in Water Channel Membranes: Lessons Learned From Polymeric Membranes and CNTs. *Faraday Discuss.* **2018**, *209* (0), 371-388. DOI: 10.1039/C8FD00054A.
(59) Green, Y.; Shloush, S.; Yossifon, G. Effect of Geometry on Concentration Polarization in Realistic Heterogeneous Permselective Systems. *Phys. Rev. E* **2014**, *89* (4), 043015. DOI: 10.1103/PhysRevE.89.043015.
(60) Pizzocchero, F.; Gammelgaard, L.; Jessen, B. S.; Caridad, J. M.; Wang, L.; Hone, J.; Bøggild, P.; Booth, T. J. The Hot Pick-up Technique for Batch Assembly of van der Waals Heterostructures. *Nat. Commun.* **2016**, *7* (1), 11894. DOI: 10.1038/ncomms11894.
(61) Polster, J. W.; Aydin, F.; de Souza, J. P.; Bazant, M. Z.; Pham, T. A.; Siwy, Z. S. Rectified and Salt Concentration Dependent Wetting of Hydrophobic Nanopores. *J. Am. Chem. Soc.* **2022**, *144* (26), 11693-11705. DOI: 10.1021/jacs.2c03436.
(62) Thompson, A. P.; Aktulga, H. M.; Berger, R.; Bolintineanu, D. S.; Brown, W. M.; Crozier, P. S.; in 't Veld, P. J.; Kohlmeyer, A.; Moore, S. G.; Nguyen, T. D.; et al. LAMMPS - a Flexible

Simulation Tool for Particle-Based Materials Modeling at the Atomic, Meso, and Continuum Scales. *Comp. Phys. Comm.* **2022**, *271*, 108171. DOI: https://doi.org/10.1016/j.cpc.2021.108171.
(63) Won, C. Y.; Aluru, N. R. Water Permeation Through a Subnanometer Boron Nitride Nanotube. *J. Am. Chem. Soc.* **2007**, *129* (10), 2748-2749. DOI: 10.1021/ja0687318.
(64) Jorgensen, W. L.; Maxwell, D. S.; Tirado-Rives, J. Development and Testing of the OPLS All-Atom Force Field on Conformational Energetics and Properties of Organic Liquids. *J. Am. Chem. Soc.* **1996**, *118* (45), 11225-11236. DOI: 10.1021/ja9621760.
(65) Berendsen, H. J. C.; Postma, J. P. M.; van Gunsteren, W. F.; DiNola, A.; Haak, J. R. Molecular Dynamics with Coupling to an External Bath. *J. Chem. Phys.* **1984**, *81* (8), 3684-3690. DOI: 10.1063/1.448118.
(66) Nosé, S. A Unified Formulation of the Constant Temperature Molecular Dynamics Methods. *J. Chem. Phys.* **1984**, *81* (1), 511-519. DOI: 10.1063/1.447334.
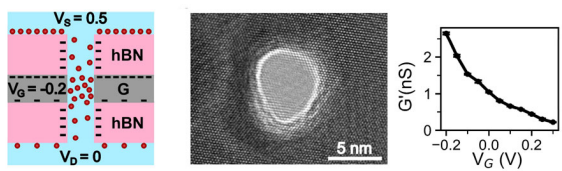

21